\documentclass[10pt,conference,twocolumn]{IEEEtran}
\usepackage{epsfig,endnotes}
\usepackage{colortbl}
\usepackage{xcolor}
\usepackage[normalem]{ulem}
\usepackage{xspace}
\usepackage{latexsym}
\usepackage{graphicx}
\usepackage[hyphens]{url} 
\usepackage{verbatim} 
\usepackage{multirow}
\usepackage{caption}
\usepackage{authblk}
\usepackage[export]{adjustbox}

\usepackage[nameinlink]{cleveref}
\crefname{section}{\S}{\S\S}
 \usepackage{amssymb}
\usepackage{cleveref}
\crefname{section}{�}{��}
\Crefname{section}{�}{��}

\usepackage{fancyhdr}
\usepackage{lipsum}
\newcommand{\tripledagger}{\ensuremath{\S}}

\begin{document}



\title{Revisiting Computational Storage for Data Integrity and Security}

    
\author{\normalsize
    {\rm Chao Shi$^\dagger$, Anthony Manschula$^\dagger$, Tabassum Mahmud$^\dagger$, Zeren Yang$^\tripledagger$,  Mai Zheng$^\dagger$} \\ Yong Chen$^\ddagger$, Jim Wayda$^\ddagger$, Matthew Wolf$^\ddagger$, Byungwoo Bang$^\ddagger$\\
    {\rm $^\dagger$Iowa State University \hspace{3em} $^\tripledagger$University of Wisconsin-Madison 
    \hspace{3em} $^\ddagger$Samsung }
}

\maketitle


\section{Introduction}
\label{sec:intro}

The idea of computational storage device (CSD) has come a long way since at least 1990s~\cite{Acharya-ActiveDisk-ASPLOS98,Riedel-ActiveStorage-VLDB98}. By embedding computing resources within  storage devices, CSDs could potentially offload computational tasks from  CPUs and enable  near-data processing (NDP), reducing data movements and/or energy consumption significantly. While the initial hard-disk-based CSDs suffer from severe limitations in terms of on-drive resources,  programmability, etc., the storage market has witnessed the commercialization of solid-state-drive (SSD) based CSDs  (e.g., Samsung SmartSSD~\cite{samsung-smartssd}, ScaleFlux CSDs~\cite{scaleflux}) recently,
which has enabled CSD-based optimizations  for a variety of  application scenarios
(e.g., ~\cite{Cao-POLARDB-FAST20,Qiao-BTreeLSMTree-FAST22,Samsung-MDPI-Electronics}). 

Nevertheless, existing CSD research efforts mainly focus on performance acceleration of regular operations, leaving the potentials  on system reliability/security largely unexplored.
In this work, we attempt to bridge the gap. We  revisit the classic idea of CSDs from a new angle: Can we leverage CSD to improve data integrity and/or  security? 
To answer the question, we look into three representative I/O-intensive reliability/security techniques for data protection, and explore their similarities and potentials for CSD-based optimizations:
\begin{itemize}
    \item \textit{Fault Injection (FI)} is an indispensable  method for testing the failure recovery of various  storage systems (e.g., ~\cite{Remzi-IRON-FileSystem-SOSP05,Om-FAST18-RFSCK,Om-TOS18,Zheng-OSDI14-DB,Remzi-FAST17-RedundencyDoesNotImplyFaultTolerance,JinruiCao-ICS18-PFault,Runzhou-TOS22-StudyPFS,hotstorage24}).
    We observe that the core operations of FI typically involve \textit{intercepting  I/O blocks} 
    at certain software layer (e.g., kernel block layer~\cite{Remzi-IRON-FileSystem-SOSP05}, FUSE~\cite{Remzi-FAST17-RedundencyDoesNotImplyFaultTolerance},  drivers~\cite{Om-TOS18,JinruiCao-ICS18-PFault,Runzhou-TOS22-StudyPFS}) to implement the functionality. A CSD-based FI solution could potentially achieve similar I/O interception and manipulation at the bottom of the storage stack (i.e., device) to enable full-stack testing with high fidelity.
    
\item  \textit{Erasure Coding (EC)} 
 is an essential fault-tolerance mechanism  for modern distributed storage systems (DSS) (e.g., Ceph~\cite{ceph}, HDFS~\cite{hdfs_msst10}).  
We observe that the core operations of EC involve \textit{matrix multiplications} for encoding/decoding. In particular, 
locally repairable codes (LRC) ~\cite{lrc-SaurabhKadekodi-fast23}
have been proposed to reduce the network and/or storage I/O cost by leveraging local parities, which could potentially benefit from FPGA-based optimization with a small set of collaborative CSDs.

\item  \textit{Ransomware Detection \& Recovery (RDR)} is increasingly important for protecting user data as ransomware has grown
to a national security threat recently~\cite{ransomtag-ccs23}. We observe that one major category of RDR solutions rely on SSDs~\cite{ssdassisted,ssdinsider,flushguard,contentbased,ransomblocker,rssd,mimosaftl}  or hypervisor~\cite{ransomtab-ccs23} to achieve \textit{I/O pattern monitoring} for ransomware detection and \textit{intra-device data movement} for data recovery, both of which aligns well with CSD  characteristics.
A CSD-based RDR 
could potentially achieve higher flexibility (compared to regular SSD-based RDR) and  efficiency (compared to hypervisor-based RDR).

\end{itemize}

Based on the key observations above, we design a generic SmartSSD CSD library called \texttt{CSDGuard} to serve as a building block for constructing CSD-optimized reliability/security solutions. 
The library follows the Computational Storage Architecture Programming Model\cite{standard}  to cover the core operations (e.g., host-device buffer management, I/O interception and monitoring, multi-dimensional array multiplication) of representative FI, EC, and RDR algorithms. Moreover, it provides a simple set of APIs to abstract away unnecessary CSD internals and support controlling data and metadata operations between host and CSDs with flexible configurability. 

To demonstrate the potential of such a solution, we build a prototype of \texttt{CSDGuard} based on the Samsung SmartSSD platform~\cite{samsung-smartssd}.
The prototype leverages the peer-to-peer (P2P) transfer between NVMe flash storage and on-drive FPGA to minimize data communication between the host and the CSD, and applies a set of 
directive-based optimizations (e.g., \texttt{HLS INTERFACE}, \texttt{HLS ARRAY\_PARTITION}, \texttt{HLS UNROLL}) to make full use of the massive parallelism of FPGA and thus achieve  efficient near data processing.
Our preliminary results are promising: 
Measuring the execution time of our library with directive-based optimizations applied, the overall latency was successfully reduced up to 70\% across several experimental data sizes (e.g., the tested matrix size ranges from 384x384 to 2048x2048). 
With regard to P2P data transfer time, we observed similar performance to the conventional software-based data transfer approach between the CSD and host device. We believe we may be incurring some additional overhead in the system calls, which may lead to the behavior that we observed.
We plan to extend the preliminary prototype to cover different use cases (e.g., FI, EC, and RDR) and evaluate with realistic systems (e.g.,  Ceph/HDFS with EC configuration) and datasets (e.g., VirusTotal) to fully demonstrate the potentials of CSD for data protection.

In order to compare the effectiveness of our design to a traditional CPU-based approach, the team collected information on data transfer and multiplication algorithm execution times. The test setup for both the CPU and SmartSSD-based approaches are as follows: For both approaches, two input square matrices of a user-defined size are randomly generated on the host device and written to the NVMe flash on the SmartSSD. Figure 1 shows the flow of data when using the SmartSSD-based approach.

\begin{figure}[ht]
\centering
\includegraphics[scale=0.15]{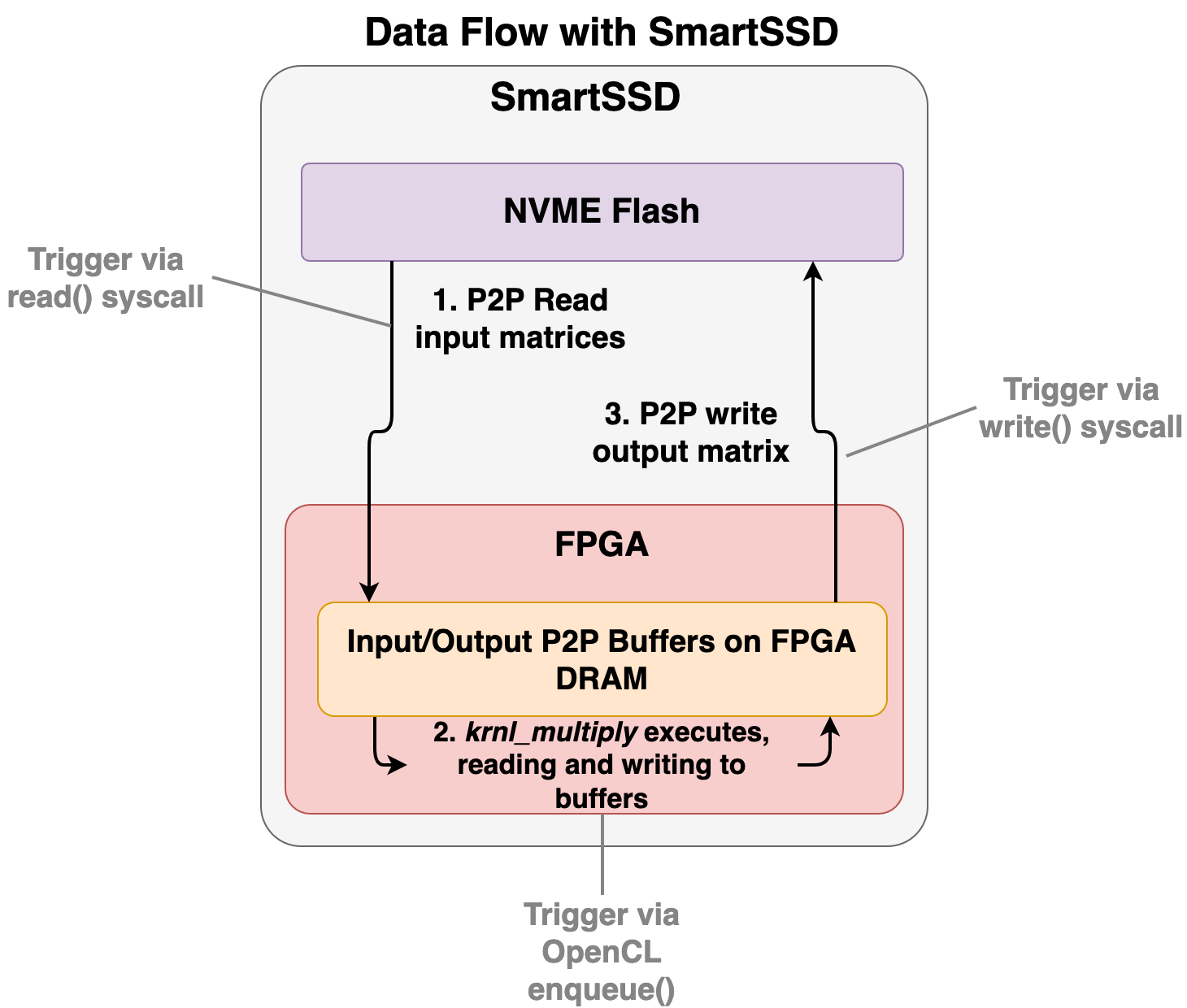}
\caption{\textit{Operation of the Samsung SmartSSD-based approach with P2P.}}
\label{fig2}
\end{figure}

For the SmartSSD-based approach, performance measurement was done via two methods. As it can be seen in Figure 2, while data movement is entirely contained on the SmartSSD, the process still needs to be initiated by a read or write system call from the host device. In order to obtain the amount of time required for the data transfers to complete, the \verb|high_resolution_clock()| function within the C++ \textit{chrono} library was utilized, taking the current time before the read/write function was called, the current time after the call returns, and subtracting the two. To obtain accelerator kernel execution time, OpenCL Events were used. The \verb|getProfilingInfo()| function provided by the Event class allows us the ability to extract performance profiling information from tasks that are queued, such as a hardware kernel. In this instance, we utilize the kernel start and end properties, subtracting in a manner as with the data transfer times. 

The CPU-based approach follows a similar methodology, with the exception of the processing kernel. For a one-to-one comparison, the C code used to synthesize the hardware kernel was directly ported to the host device. The matrix computation code remained identical, with minor modifications (not impacting functionality) to read/write to the local buffers allocated on host memory as opposed to the FPGA DRAM. Performance measurement for the CPU-based solution was performed in a similar fashion to the SmartSSD-based solution. In this instance, we exclusively use the C++ \textit{chrono} library to measure start and end times of data transfers and matrix processing. The behavior of the CPU-based system is outlined in Figure 2 below.

\begin{figure}[ht]
\centering
\includegraphics[scale=0.165]{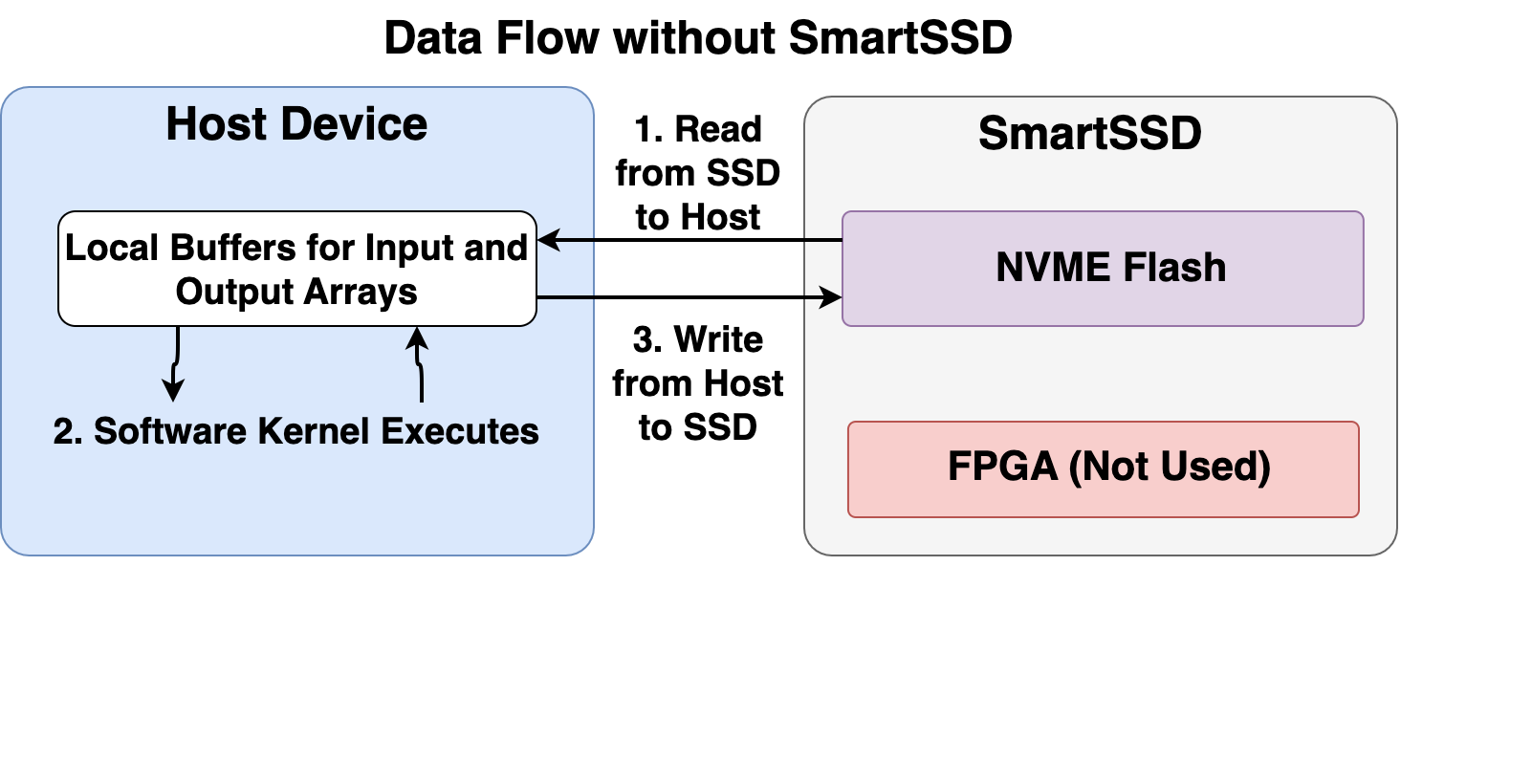}
\caption{\textit{Operation of the system using the CPU-based approach.}}
\label{fig3}
\end{figure}

The system was tested with three different data input matrix sizes - 576KiB, 4MiB, and 9.2MiB. The first of the three sizes was chosen for a point of comparison with the unoptimized design, as that was the previous maximum size the kernel could handle. 4MiB (1024x1024 unsigned integer matrix) was chosen as an intermediate size during further testing in anticipation of the next size being a 16MiB (2048x2048 matrix). However, for reasons discussed in section 3.3, this was not possible, so we landed on a maximum of about 9.2MiB, or a 1536x1536 matrix. To ensure consistency in the execution time data that was collected for both approaches, 2000 consecutive runs of data reading and writing were made each test and 50 consecutive kernel runs were performed. The reasoning behind the lower number of kernel repetitions versus read/write repetitions is that the processing aspect exhibited much less run-to-run inconsistency when only one repetition was performed, allowing us to save time testing while still giving us confidence that the result was accurate.

\begin{figure}[ht]
\centering
\includegraphics[scale=0.4]{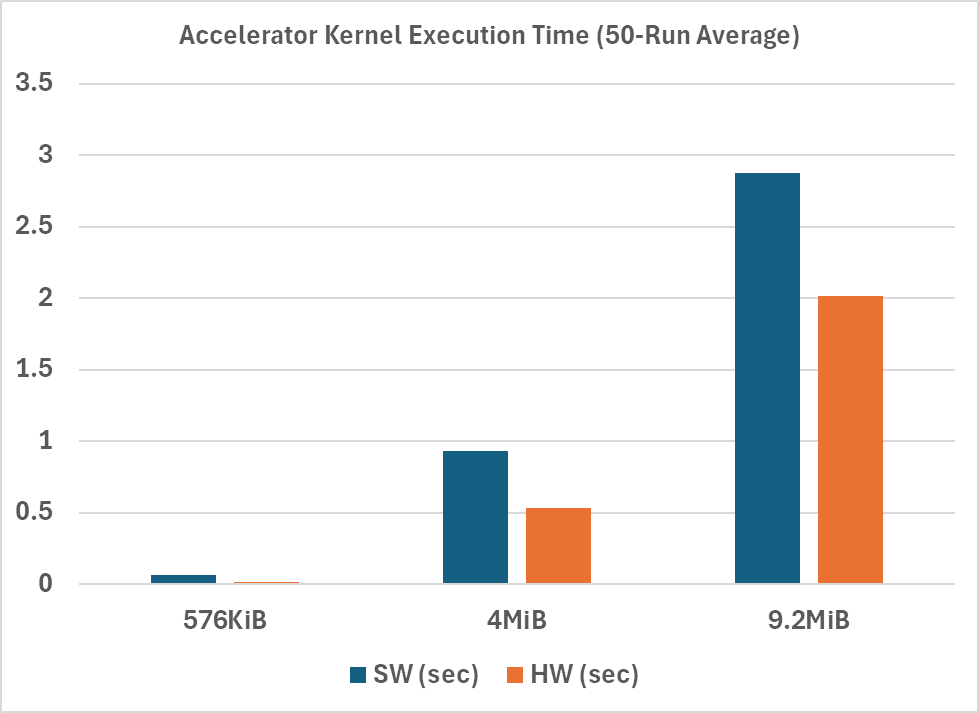}
\caption{\textit{Kernel processing time for each data size, measured in seconds.}}
\label{fig4}
\end{figure}

We observed that the execution time of both the hardware and software kernel grew nonlinearly as the input data size scaled up. The software kernel required from 0.062 seconds to complete at the 576KiB input size to as much as 2.876 seconds to complete at the 9.2MiB input size. The hardware kernel performed much better than the software kernel across the board, seeing over 3x speedup at the 576KiB level, down to about a 1.4x speedup at the 9.2MiB level. As a whole, the design showed significant improvement versus previous iterations. Compared to the midterm implementation of the accelerator, which required over 200ms to compute an output from 576KiB of input data in hardware, the new implementation completes in just 18ms. It is also worth considering that the results of the software implementation represent the performance of high-end server hardware, meaning that the FPGA implementation would likely pull further ahead when compared to a system with more pedestrian components.

\begin{figure}[ht]
\centering
\includegraphics[scale=0.4]{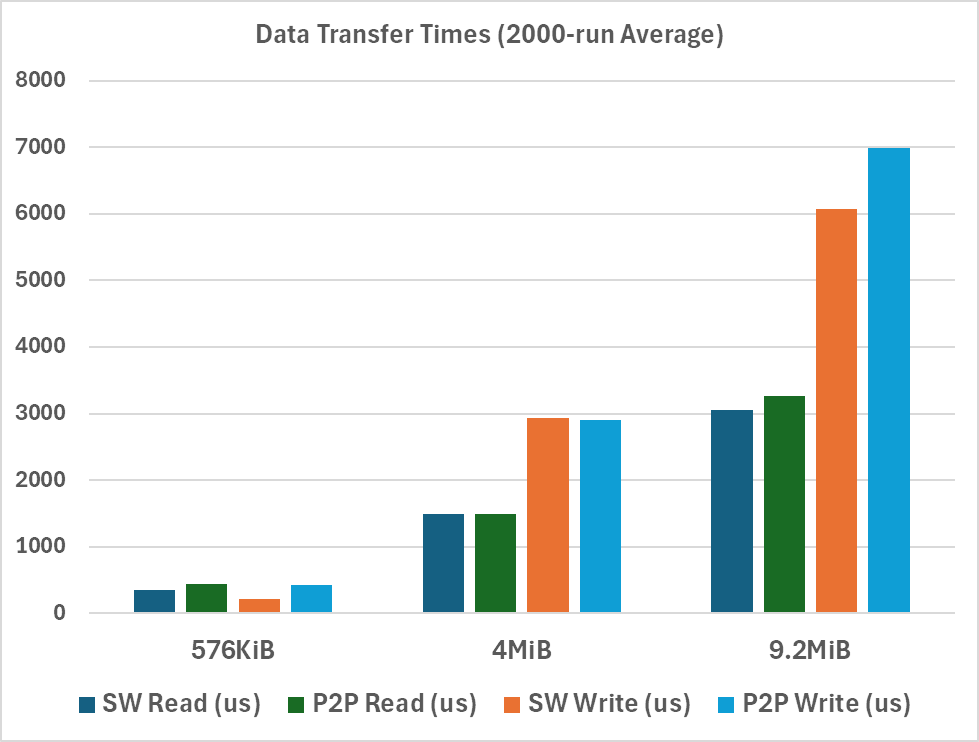}
\caption{\textit{Time to complete data transfers of 576KiB, 4MiB, and 9.2MiB, measured in microseconds.}}
\label{fig5}
\end{figure}

Data transfer times tell a slightly different story. For one, the scaling as data size increases appears to be mostly linearly correlated. Furthermore, the transfer times for the CPU-based approach and the HW-based approach are largely the same across data sizes. Additionally, the run-to-run variance was sometimes quite large - taking a 2000-run average smooths the data for the most part, however some outliers like the 9.2MiB P2P write still show odd variance. This data will be discussed in more detail in section 3.3.
\begin{figure}[ht]
\centering
\includegraphics[scale=0.4]{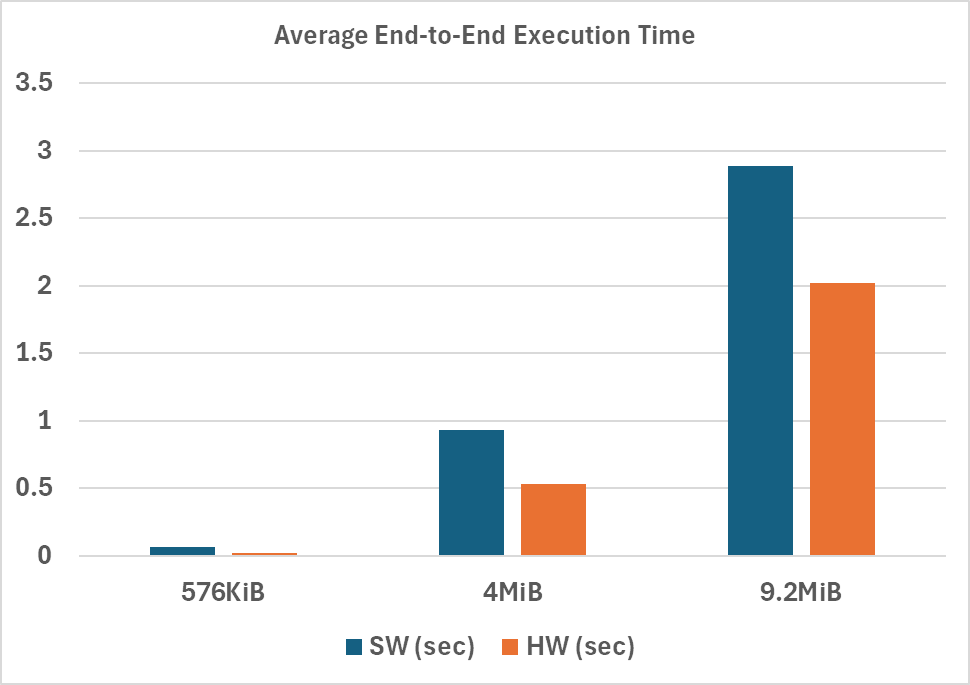}
\caption{\textit{End-to-End program execution time, including kernel and data transfer times, measured in seconds.}}
\label{fig6}
\end{figure}
The plot of end-to-end execution time of the software and hardware routines shown in Figure 6 has a nearly identical appearance to the kernel execution time chart. This is because data transfer times account for only a small portion of the overall latency, being around only a couple percent on average across all runs of hardware and software implementations.

An important aspect of the design to consider is the characteristics of the underlying hardware that is synthesized by Vitis HLS. As mentioned, the more performant version of our accelerator design utilizes loop unrolling; Loop unrolling creates many copies of the loop body for different iterations of the loop, allowing them to execute concurrently and (in theory) reducing the amount of time spent processing the loop. However, a critical downside to such an approach is that the complexity and physical size of the synthesized design increases substantially, which has a negative impact on maximum clock speed. The team observed this effect firsthand: Kernels with larger input array sizes (meaning more loop iterations to fully compute the output matrix) led to a substantial reduction in final clock speed (see Figure 7). This reduction in clock speed likely contributes to the HW solution's observed decline from a 3x advantage to only 1.4x.

\begin{figure}[ht]
\centering
\includegraphics[scale=0.4]{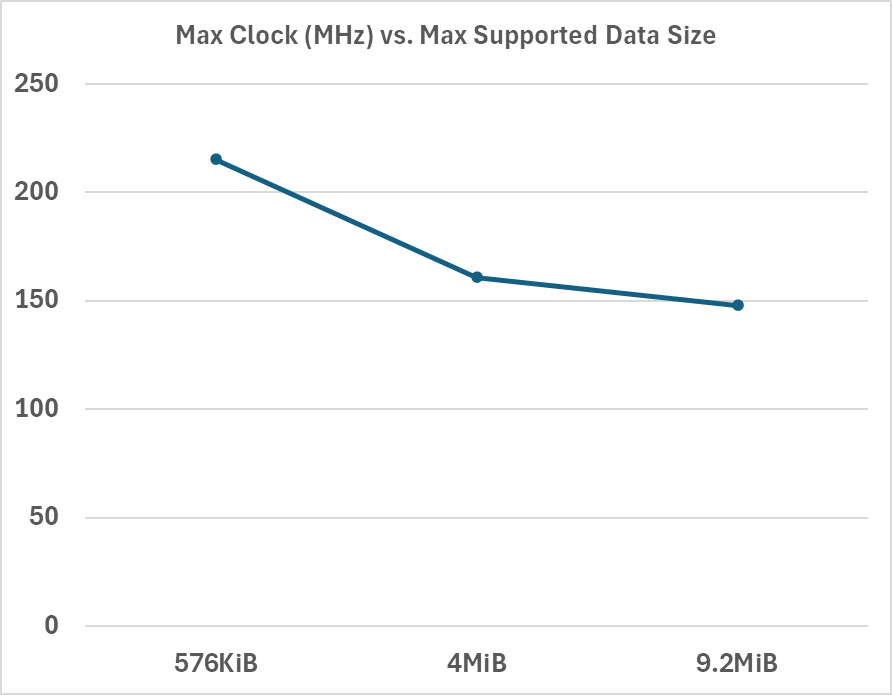}
\caption{\textit{Max achievable clock speed compared to maximum supported input data size, as reported by the synthesis tool.}}
\label{fig7}
\end{figure}

\noindent
\textbf{Acknowledgements}:
This work was supported in part by National Science Foundation (NSF) under grands CNS-1855565 and CNS-1943204, and a Global Research Outreach (GRO) Award from Samsung (2022).




\end{document}